# Title: An algebraic theory to discriminate qualia in the brain

**Yoshiyuki Ohmura**[*,1], **Wataru Shimaya**[1] **& Yasuo Kuniyoshi**[1]

[1] *Department of Mechano-Informatics, Graduate School of Information Science and Technology, The University of Tokyo, Tokyo 113-8656, Japan*

Email: {ohmura, shimaya, kuniyosh}@isi.imi.i.u-tokyo.ac.jp

**Corresponding author:** Yoshiyuki Ohmura,

Department of Mechano-Informatics, Graduate School of Information Science and Technology, The University of Tokyo, 7-3-1 Hongo, Bunkyo-ku, Tokyo, Japan, 113-8656

Phone: +81-3-5841-6339

Email: ohmura@isi.imi.i.u-tokyo.ac.jp

**Abstract**

**The mind-brain problem is to bridge relations between in higher-level mental events and in lower-level neural events. To address this, some mathematical models have been proposed to explain how the brain can represent the discriminative structure of qualia, but they remain unresolved due to a lack of validation methods. To understand the qualia discrimination mechanism, we need to ask how the brain autonomously develops such a mathematical structure using the constructive approach. In unsupervised representation learning, independence between axes is generally used to constrain the latent vector but independence between axes cannot explain qualia type discrimination because independent axes cannot distinguish between inter-qualia type independence (e.g., vision and touch) and intra-qualia type independence (e.g., green and red). We hypothesised that inter-axis independence must be weakened in order to discriminate qualia types. To solve the problem, we formulate an algebraic independence to link it to the other-qualia-type invariant transformations, whose transformation value is a vector space rather than a scalar. In addition, we show that a brain model that learns to satisfy the algebraic independence between neural networks separates metric spaces corresponding to qualia types. Learning algebraic independence is a general method for constructing independent metric spaces without constraining the axes in each space to be independent.**

**Introduction**

How can the brain discriminate different qualities of perception (qualia)? In neuroscience, differences in qualia are reduced to the differences in different domains such as brain region and pathway (Müller 2017), attractor in dynamics (Freemann 1987), information processing (Jackendoff 1987) and sensory-motor loop (O'Regan and Noë 2001). These reductions only provide the correlations between two domains, they cannot explain the mechanism to discriminate qualia in the brain. A mathematical formulation to explain the differences in the qualia is essential to understand such a mechanism.

The best formulation of the structure of qualia is currently under debate (Stanley 1999; Fekete and Edelman 2011; Yoshimi 2011; Prentner 2019; Northoff *et al.* 2019; Signorelli *et al.* 2021). These studies aimed to provide a mathematical account of the structure of similarities and differences in qualia, but the best is not established due to a lack of testing or comparison methods (Shalmers 1996; Seth and Bayne 2022; Del Pin *et al.* 2021; Browning and Veit 2020; Pinto and Stein 2021). A speculative account of the properties of qualia is not sufficient to test the theories. To clarify the mechanism that discriminates qualia in the brain, we need to ask how the brain autonomously develops such a mathematical structure. Unsupervised learning is necessary because the subjective perception must be explained by one's own ability (Fekete and Edelman 2011; Piaget 1950; Husserl 1958; Graziano 2022) rather than by the derivatives from the outside (Searle 1980). Using a constructive method (Kuniyoshi 2018; Chrisley 2008; Clowes and Seth 2008; Seth 2009) to reproduce the development of the structure of qualia provides an opportunity for testing. However, qualia discrimination cannot currently be achieved by fully unsupervised learning (Fumero *et al.* 2021). Furthermore, evolutionary learning cannot currently explain how the subjective perception represents the properties of physical object models because it is tuned to fitness rather than truth in world model (Prakash *et al.* 2020).

We considered that different qualia types have different corresponding metric spaces, because it is difficult to judge the similarity between qualia of different types, but it is possible within the same type (Jackendoff 1987; O'Regan and Noë 2001). For example, we can notice that the experience of red is more similar to the experience of pink than to that of black (O'Regan and Noë 2001). But we cannot judge whether the experience of red is more like the experience of a triangle than that of a square. Furthermore, the absolute values of colour, shape, size or position in experience are not comparable inter-individuals or inter-species (Nagel 1974; Kleiner 2020), although their relative values can be compared through verbal reports or behavioural responses. These properties suggest that different types of qualia have different corresponding metric spaces in each individual. The remaining issue is how to formulate differences between metric spaces that can explain the natural distinction between qualia.

In this study, we show that the brain can autonomously construct different metric spaces corresponding to each qualia type by learning to satisfy an algebraic independence (Simpson 2018) between neural network modules.

*Why focus on algebra between neural networks?*

Descartes argued that the mind is not matter. Current neuroscientists believe that the mind is caused by the brain. The problem is to clarify how the mind, which cannot be reduced to the brain and its activity, has unique causal power over the brain (Sperry 1987; Van Gulick 2001; Chalmers 2006; Campbell 1974) without any additional matter other than the brain. To solve this problem, we focus on the macro-micro hierarchy (Simon 1977): the macro-components consist of micro-elements, and they don't need any additional matter other than the micro-elements. In the brain, the macro-components are neural networks (i.e. functions). Then a relationship between the macro-components is an algebra that cannot be reduced to the law of the micro-elements. We hypothesised that an algebra between neural networks can change the activity of neurons. We noticed that learning algebra between neural networks is a mathematical formulation of the downward causation (Campbell 1974; Sperry 1987; Van Gulick 2001; Chalmers 2006). To our knowledge, only Piaget focused on developmental change in algebraic structure of intelligence (Piaget 1950). Dualism of algebra and matter opens up the possibility of a causal mechanism from the mind to the body.

*A functional role of the downward causation from algebra to the brain*

What is the functional role of the downward causation from algebra to the brain? We focused on the algebraic independence (Simpson 2018) which is a generalization of several mathematical independences including orthogonality and stochastic independence because Stanley argued that a new definition of orthogonality is required to represent qualia space mathematically (Stanley 1999).

In representation learning (Fumero *et al.* 2021; Higgins *et al.* 2017; Comon 1994), independence between axes rather than spaces is often used to constrain the latent vector, but the constraint cannot account for qualia type discrimination because independence between axes cannot distinguish between inter-qualia type independence (e.g., vision and sound) and intra-qualia type independence (e.g., red and green). To discriminate qualia type, independence between metric spaces whose axes are not mutually independent has to be defined.

The independence between the axes can be interpreted as an invariant transformation of a scalar transformation parameter. The invariant transformation is a natural formulation of the permanence of perception (Pitts and McCulloch 1947; Cassirer 1944; Hoffman 1966) and it is used for the theory of pattern recognition (Otsu 1986) and the representation learning (Higgins *et al.* 2018). For example, shape qualia are invariant to position and rotation transformations. And the scalar parameter invariant transformation is an intuitive interpretation of orthogonality.

In this article, we show that the algebraic independence between transformations is the minimum condition to define a generalized invariant transformation of a vector transformation parameter. In addition, the vector transformation parameter can be interpreted as a metric space corresponding to each qualia type.

*The invariant transformation equation satisfies an algebraic independence structure*

The algebraic independence structure in category theory (Simpson 2018) is a generalization of several independences in mathematics including the stochastic independence used in representation learning (Fumero *et al.* 2021; Higgins *et al.* 2017). Here, based on this algebraic independence, we newly formulate an algebraic independence between transformations to show a connection to the invariant transformation equation. In the formulation, we limit the number of transformations to two for simplicity, but a formulation allowing arbitrary number of transformations can be extended from the present formulation (Supplementary Information).

We define observation space as a countable set $S$ of $\mathbb{R}^N$. Let a relationship between two points $X, Y \in S$ be denoted as $Y = F_0(\lambda_0)F_1(\lambda_1)[X]$, where $F_0$ and $F_1$ are transformations (functions whose input and output dimensions are the same) and $\lambda_0 \in \mathbb{R}^{n_0}$ and $\lambda_1 \in \mathbb{R}^{n_1}$ are transformation parameters. Both transformation parameters are vector spaces. If the two transformations, $F_0(\lambda_0)$ and $F_1(\lambda_1)$, satisfy the following conditions, they are algebraically independent.

Condition 1 is the existence of unit element, $\exists \lambda_0^I : F_0(\lambda_0^I) = I$ and $\exists \lambda_1^I : F_1(\lambda_1^I) = I$, where $I$ is the identity transformation. Hereafter, the unit transformation parameters will be denoted as $\mathbf{0}$. From condition 1, we have an invariant condition of the transformation such as $\forall \lambda_0 : F_0(\lambda_0)F_0(\mathbf{0}) = F_0(\lambda_0)$.

Condition 2 is the commutativity, $F_0(\lambda_0)F_1(\lambda_1) = F_1(\lambda_1)F_0(\lambda_0)$. From condition 2, the unique composition of two transformations, Y, can be obtained from X and transformation parameters, $\lambda_0$ and $\lambda_1$.

Condition 3 is the uniqueness of transformation parameters. $Y = F_0(\lambda_0)F_1(\lambda_1)[X] = F_1(\lambda_1)F_0(\lambda_0)[X] = \Phi(\lambda_0, \lambda_1, X)$, where $\Phi$ is an bijective function. In other words, the transformation parameters $\lambda_0$ and $\lambda_1$ are uniquely determined from X and Y.

When we change the transformation parameter $\lambda_0$ to $\lambda_0'$, we uniquely obtain $Y' = \Phi(\lambda_0', \lambda_1, X)$. Therefore, the transformation $F_0$ is a $\lambda_1$-invariant transformation. Similarly, the transformation $F_1$ is a $\lambda_0$-invariant transformation. Thus, the algebraic independence between transformations defines the invariant transformation equation.

*Projection to multiple metric spaces*

We constructed a projection from the distance between X and Y to two norms $\|\lambda_0\|$ and $\|\lambda_1\|$. Therefore, the points, X and Y, project to two latent spaces, $Q_0 \subset \mathbb{R}^{n_0}$ and $Q_1 \subset \mathbb{R}^{n_1}$, respectively. Then, we defined the transformation parameters, $\lambda_0$ and $\lambda_1$, by subtraction of latent vectors corresponding to X and Y. So, we used a familiar encoder-decoder model in representation learning (Fumero *et al.* 2021; Higgins *et al.* 2017). Both encoder and decoder are often implemented using neural networks. Theorem 1 provides the conditions to satisfy the algebraic independence between transformations, $F_0$ and $F_1$, when the corresponding transformations projected in the latent space, $f_0$ and $f_1$, satisfy the algebraic independence.

**Theorem 1 (projection of transformations)**

Given $F_0(\lambda_0) = G_N f_0(\lambda_0) G_P$ and $F_1(\lambda_1) = G_N f_1(\lambda_1) G_P$, where $G_P$ is an encoder and $G_N$ is an bijective decoder. If transformations, $f_0$ and $f_1$, in the latent spaces satisfy the algebraic independence and the following $G_P - F$ commutativity holds, then the corresponding transformations, $F_0$ and $F_1$, also satisfy the algebraic independence.

$G_P - F$ commutativity:
$$G_P F_0(\lambda_0)[X] = G_P G_N f_0(\lambda_0) G_P[X] = f_0(\lambda_0) G_P[X]$$
$$G_P F_1(\lambda_1)[X] = G_P G_N f_1(\lambda_1) G_P[X] = f_1(\lambda_1) G_P[X]$$

**Proof**

Commutativity is obtained from the $G_P - F$ commutativity and the above definitions as follows:
$$F_0(\lambda_0) F_1(\lambda_1)[X] = G_N f_0(\lambda_0) G_P G_N f_1(\lambda_1) G_P[X] = G_N f_0(\lambda_0) f_1(\lambda_1) G_P[X] = G_N f_1(\lambda_1) f_0(\lambda_0) G_P[X]$$
$$= G_N f_1(\lambda_1) G_P G_N f_0(\lambda_0) G_P[X] = F_1(\lambda_1) F_0(\lambda_0)[X]$$

Similarly, the existence of unit element is obtained because $f_0$ and $f_1$ satisfy the condition 1 for algebraic independence:
$$F_0(\mathbf{0}) F_1(\lambda_1)[X] = G_N f_0(\mathbf{0}) G_P G_N f_1(\lambda_1) G_P[X] = G_N G_P G_N f_1(\lambda_1) G_P[X] = G_N f_1(\lambda_1) G_P[X]$$
$$= F_1(\lambda_1)[X]$$
$$F_1(\mathbf{0}) F_0(\lambda_0)[X] = G_N f_1(\mathbf{0}) G_P G_N f_0(\lambda_0) G_P[X] = G_N G_P G_N f_0(\lambda_0) G_P[X] = G_N f_0(\lambda_0) G_P X$$
$$= F_0(\lambda_0)[X]$$

The uniqueness of the transformation parameters is obtained from the $G_P - F$ commutativity and the above definitions as follows:

If we defined the bijective function $\varphi$ as $\varphi(\lambda_0, \lambda_1, X) = f_0(\lambda_0) f_1(\lambda_1) G_P[X]$, then $\Phi(\lambda_0, \lambda_1, X) = F_0(\lambda_0) F_1(\lambda_1)[X] = G_N f_0(\lambda_0) G_P G_N f_1(\lambda_1) G_P[X] = G_N f_0(\lambda_0) f_1(\lambda_1) G_P[X] = G_N \varphi(\lambda_0, \lambda_1, X)$, where $\Phi$ is the bijective function because both $G_N$ and $\varphi$ are bijective. This way we obtained the desired result.

Next, we designed the transformations in the latent space that would allow the algebraic independence to hold. We defined the latent vectors $\mathbf{x}$ and $\mathbf{y}$ on $\mathbb{R}^{n_0+n_1}$ as follows:

$$\mathbf{x} = (\mathbf{x}_0, \mathbf{x}_1) = (G_{P0}[X], G_{P1}[X]) = G_P[X]$$
$$\mathbf{y} = (\mathbf{y}_0, \mathbf{y}_1) = (G_{P0}[Y], G_{P1}[Y]) = G_P[Y]$$
$$\mathbf{x}_0, \mathbf{y}_0 \in Q_0$$
$$\mathbf{x}_1, \mathbf{y}_1 \in Q_1$$

Consider the transformations $f_0(\lambda_0): \mathbf{x}_0 \to \mathbf{y}_0$ and $f_1(\lambda_1): \mathbf{x}_1 \to \mathbf{y}_1$, which are transformations on different latent spaces. These transformations satisfy the algebraic independence. Then, the $G_P - F$ commutativity can be rewritten as $(\mathbf{y}_0', \mathbf{x}_1') \stackrel{\text{def}}{=} G_P G_N(\mathbf{y}_0, \mathbf{x}_1) = (\mathbf{y}_0, \mathbf{x}_1)$ and $(\mathbf{x}_0', \mathbf{y}_1') \stackrel{\text{def}}{=} G_P G_N(\mathbf{x}_0, \mathbf{y}_1) = (\mathbf{x}_0, \mathbf{y}_1)$ because $f_0(\lambda_0) G_P[X] = (\mathbf{y}_0, \mathbf{x}_1)$ and $f_1(\lambda_1) G_P[X] = (\mathbf{x}_0, \mathbf{y}_1)$ by definitions.

To reconstruct the observation Y from the latent vectors, $Y = G_N(\mathbf{y}_0, \mathbf{y}_1)$ needs to be maintained. Combining the reconstruction condition and $G_P - F$ commutativity, we obtain the following loss functions, $\|Y - G_N(\mathbf{y}_0, \mathbf{y}_1')\|$ and $\|Y - G_N(\mathbf{y}_0', \mathbf{y}_1)\|$. Figure 1 shows a summary of our encoder-decoder model. If $Y = G_N(\mathbf{y}_0, \mathbf{y}_1') = G_N(\mathbf{y}_0', \mathbf{y}_1)$, then $\mathbf{y}_0 = \mathbf{y}_0'$ and $\mathbf{y}_1 = \mathbf{y}_1'$ in the $G_P - F$ commutativity hold because $G_N$ is the bijective function. The same is true for X.

After $G_P$ and $G_N$ learning to satisfy the loss function, we obtained two distances between X and Y from the projections as follows:

$$\|Y - X\|_0 \stackrel{\text{def}}{=} \|G_{P0}Y - G_{P0}X\| = \|\mathbf{y}_0 - \mathbf{x}_0\|$$
$$\|Y - X\|_1 \stackrel{\text{def}}{=} \|G_{P1}Y - G_{P1}X\| = \|\mathbf{y}_1 - \mathbf{x}_1\|$$

In this theory, both transformations, $f_0$ and $f_1$, in the latent spaces are maps between two latent vectors projected from the countable set $S$. Therefore, the transformations are neither a continuous group nor the Lie group of a scalar transformation parameter, which are often used in invariant transformation theories (Hoffman 1966; Otsu 1986).

In representation learning, all axes in the latent vector are often constrained to be mutually orthogonal or stochastically independent. However, this constraint cannot account for qualia type discrimination. Therefore, the constraint has to be weakened from between axes to between spaces, and the axes of the spaces should not be constrained to be independent.

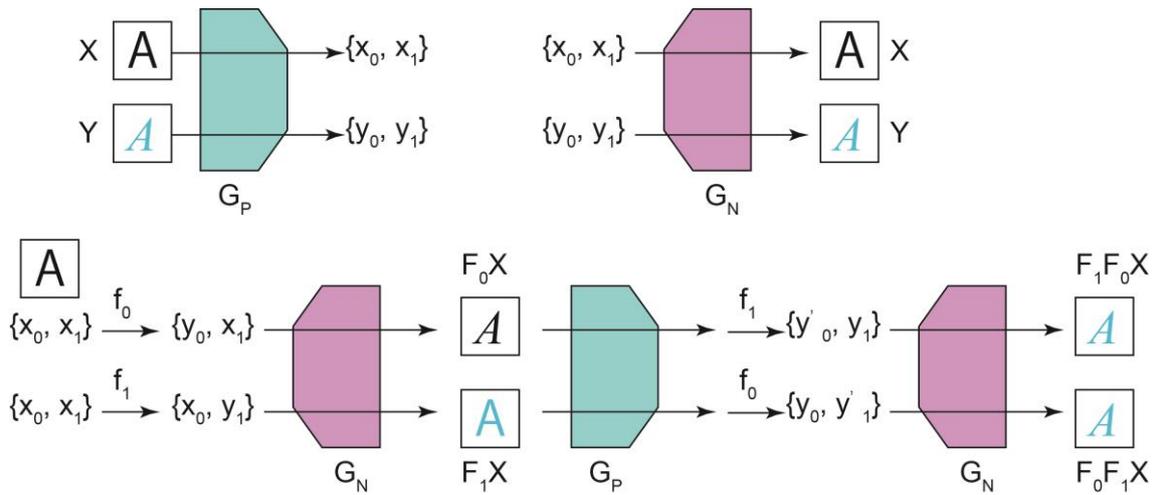

Figure 1: Information flow. The network consists of two networks, $G_P$ and $G_N$, where $G_P$ is an encoder from sensory input to latent space and $G_N$ is a decoder from latent space to sensory input. Transformations $f_0$ and $f_1$ in the latent space transform the latent vector x into the latent vector y.

*Embedding into the neural network model*

We verified that the brain model which would learn the algebraic independence structure between neural networks could represent the invariant transformation between sensory information and construct multiple metric spaces. We assumed $G_{P0}$, $G_{P1}$ and $G_N$ in the above formulation were neural networks with network parameters trained to satisfy the $G_P - F$ commutativity. Then, we used the globally injective ReLU network (Puthawala *et al.* 2022) for the bijective function, $G_N$. To satisfy the $G_P - F$ commutativity, we used the following loss function; $\|Y - G_N(y_0, y_1')\| + \|Y - G_N(y_0', y_1)\|$, where $F_0 F_1 X \stackrel{\text{def}}{=} G_N(y_0, y_1') = G_N(G_{P0}Y, G_{P1}G_N(G_{P0}X, G_{P1}Y))$ and $F_1 F_0 X \stackrel{\text{def}}{=} G_N(y_0', y_1) = G_N(G_{P0}G_N(G_{P0}Y, G_{P1}X), G_{P1}Y)$.

**Methods**

*Dataset*

Dataset consists of 26 alphabets with 12 fonts and 7 colours. Image size is 3 channels × 32 pixels × 32 pixels. The background color is black, (0,0,0) in (R, G, B). Seven colors of alphabets consist of (0, 0, 1), (0, 1, 0), …, (1, 1, 1). Two images are randomly sampled to make pairs of (X, Y). To increase the variety of colours, a random value from 0.2 to 1 was multiplied to the colour channels.

*Network structures*

$G_{P0}, G_{P1}$:

The input image is convolved using three Convolutional Neural Networks (CNN) (Krizhevsky *et al.* 2012) whose kernel size is 4, stride is 2 and padding is 1 without bias and two linear layers follow. We used ReLU function for activation function. The channels of the convolution layers were 128, 256 and 512. The output dimensions of the linear layers were 4096 and 32. The dimension of both latent vectors were 32.

$G_N$:

To realize bijective function, we used Globally Injective ReLU Network (Puthawala *et al.* 2022). The input was tuple of two latent vectors $(x_0, x_1)$. The network consists successively of three linear layers without bias and three Transposed Convolution layers whose kernel size is 4, stride is 2, and padding is 1 without bias and final layer is invertible Conv1x1 layer (Lin et al. 2014; Kigma and Dhariwal 2018). The output channels of linear layers were 128, 1024 and 4096. The Globally Injective ReLU linear layer is defined by $(Q, -Q, W, -W)$, where $Q \in R^{n_{in} \times n_{in}}$ and $W \in R^{\frac{n_{out}}{2} - n_{in} \times n_{in}}$, $n_{in}$ is the dimension of input channel and $n_{out}$ is the dimension of output channel. Q is constrained to the orthogonal matrix using Trivialization method (Lezcano-Casado 2019).

In the Transposed Convolution layer, when convolution matrix is denoted by C whose input is $(ch_{in}, width, height)$-size tensor and output is $(ch_{in}/4, width * 2, height * 2)$-size tensor, the layer is defined by $(C, -C)$. Output of Conv1x1 layer was divided into two outputs at channels. The first output is image output with $(3, \text{width}, \text{height})$. The second output N with $(\text{ch} - 3, \text{width}, \text{height})$ learns to approach to 0 by the following auxiliary loss, where ch is the order of square matrix in the weight of Conv1x1 layer. To avoid the weight of Conv1x1 converging to 0, we used the modified weight normalization (Salimans and Kingma 2016; Hoffer *et al.* 2018) whose norm of weigh in each output was fixed to initial value, for the last Conv1x1 layer. We set the ch to 32. We used the same network configuration using different initialization for comparisons.

*Loss functions*

Let points X and Y on observation space $S \in \mathbb{R}^N$. The latent vectors are calculated by two independent encoders, $G_{P0}$ and $G_{P1}$ as follows.

$$(\boldsymbol{x}_0, \boldsymbol{x}_1) = (G_{P0}X, G_{P1}X)$$
$$(\boldsymbol{y}_0, \boldsymbol{y}_1) = (G_{P0}Y, G_{P1}Y)$$

Using the single-transformation images, $(F_0X, N_0) = G_N(\boldsymbol{y}_0, \boldsymbol{x}_1)$ and $(F_1X, N_1) = G_N(\boldsymbol{x}_0, \boldsymbol{y}_1)$, we calculated the latent vectors of the constructed single-transformation images as follows.

$$\boldsymbol{y}'_0 = G_{P0}F_0X$$
$$\boldsymbol{y}'_1 = G_{P1}F_1X$$

Finally, we calculated the reconstruction images as follows.

$$(F_1F_0X, N_{10}) = G_N(\boldsymbol{y}_0, \boldsymbol{y}'_1)$$
$$(F_0F_1X, N_{01}) = G_N(\boldsymbol{y}'_0, \boldsymbol{y}_1)$$

We defined the auxiliary loss as $\text{aux}_{\text{loss}} = \|N_{10}\| + \|N_{01}\|$.

We defined the loss function as $\text{loss} = \|Y - F_1F_0X\| + \alpha\|Y - F_0F_1X\| + \beta * \text{aux}_{\text{loss}}$, where $\alpha$ and $\beta$ are hyper parameters. We set $\alpha$ to 1 in typical condition and to 0 in ablation condition. We set $\beta$ to 0.01(see also Supplementary Materials).

*Training*

We used RAdam (Liu *et al.* 2020) for optimization. The learning rate was 1e-4 and batch size was 128. We used CUDA 11.4, PyTorch 1.10.0 (Paszke *et al.* 2019) and Nvidia RTX3080Ti for training. Training epochs were 400 for comparison experiments or 1000 for other experiments.

*Evaluation*

**Colour invariance**: Each channel in the single-transformation images, $F_0X$ and $F_1X$ are binarized by threshold $\tau_c$. We set $\tau_c$=0.1. As a result, the binarized images have eight colors (0, 0, 0), (0, 0, 1), …, (1, 1, 1). We defined the most frequent colour except background colour (0, 0, 0) as letter colour. We calculate the inner product of normalized vectors between the letter colour of X and that of $F_0X$ or $F_1X$. We define the mean of the results in the batch as color invariance. The color invariance is close to 1 when the transformation result does not change the color of input image. We set batch size

to 128.

**Shape invariance**: The single-transformation images, $F_0 X$ and $F_1 X$ are binarized by threshold $\tau_s$. We set $\tau_s=0.1$. We define the mean of the inner product of normalized vectors in the batch between the binarized image of input X and the binarized image of $F_0 X$ or $F_1 X$ as shape invariance. The shape invariance is close to 1 when the transformation result does not change the shape of input image.

**Invariance**: We define a transformation with lower colour invariance than another transformation as shape transformation. And we define a transformation with lower shape invariance than another transformation as colour transformation. We define invariance as the mean of a colour invariance of the shape transformation and a shape invariance of the colour transformation.

**Results**

Our training dataset consisted of 26 alphabets with 12 fonts and seven colours. Data X and Y were randomly sampled from this dataset. To increase the variety of colours, a random value was multiplied by the colour channels of the sampled data. We obtained two images, $F_0 X$ and $F_1 X$, which were defined as follows:

$$F_0 X = G_N(G_{P0} Y, G_{P1} X)$$
$$F_1 X = G_N(G_{P0} X, G_{P1} Y)$$

These images are the result of a single-transformation applied to the input. These are expected to be single qualia-type transformations as a result of learning.

With training, two images, $F_0 X$ and $F_1 X$, became which a colour-only transformation or a shape-only transformation. We observed that the colour-only transformation changes only hue without changing the contrast (Fig. 2a).

In the ablation condition which learns only a reconstruction loss $\|Y - F_1 F_0 X\|$, single-transformations could hardly become single qualia transformations. Only seven of 100 trials learned the single qualia-type transformations. Thirty-four out of 100 trials learned an all-qualia transformation and an identity function (Fig. 2b). In these cases, the both reconstruction images, $F_0 F_1 X$ and $F_1 F_0 X$, converged to Y (Fig. 2b). Fifty-two out of 100 trials learned a collapsed single-transformation that could not reconstruct the shape of the input image (Fig. 2c). In these cases, $F_0 F_1 X$ could not converge to Y.

We evaluated the colour-invariance and shape invariance of the single-transformation (Fig. 2d-f). The invariance during training reflects the timeline of single-transformation learning (Fig. 2a-c). The invariance, which corresponds to the mean of the colour-invariance of the shape transformation and the shape invariance of the colour transformation, significantly differed between the control condition and the ablation condition after training (Fig2. g, Mann–Whitney U = 177, n = 100, P < $5 \times 10^{-32}$, two-tailed). Similarly, the $G_P - F$ commutativity loss after training significantly differed between conditions (Fig2. h, Mann–Whitney U = 8273, n = 100, P < $2 \times 10^{-15}$, two-tailed). Thus, $G_P - F$ commutativity learning is required to stably categorize the colour and shape transformation.

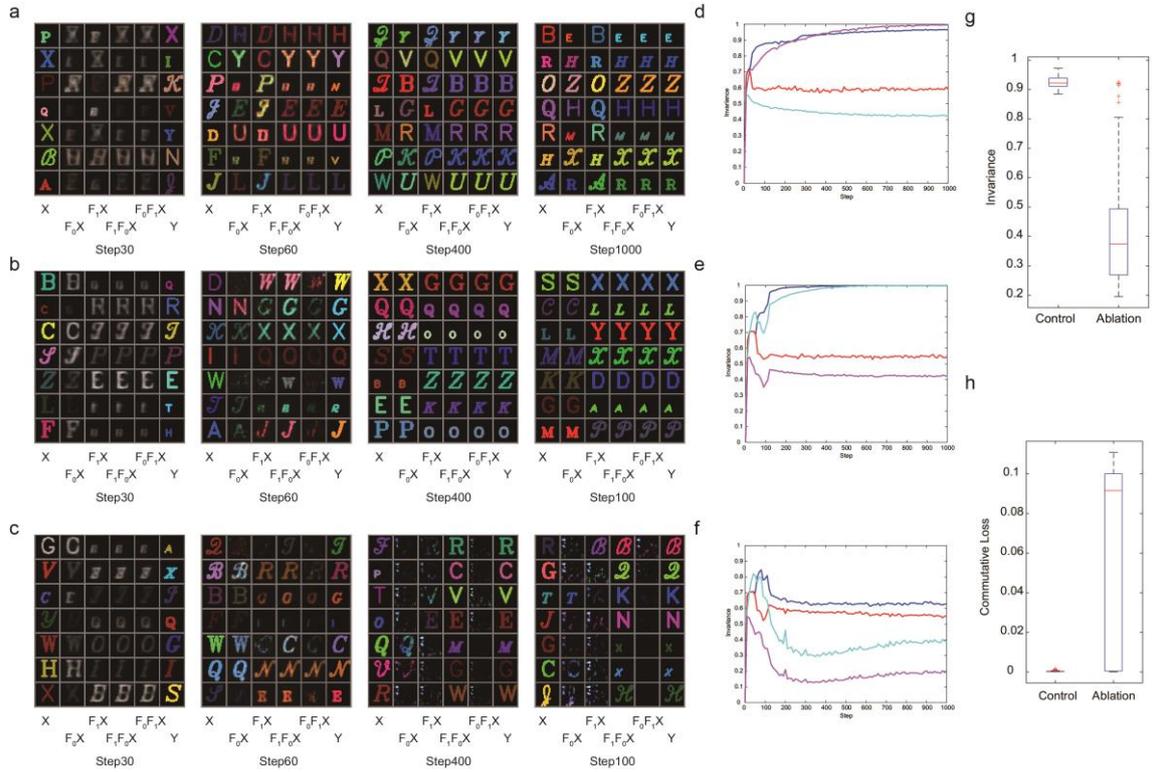

Figure 2: A brief timeline of algebraic independence learning. a, A result with commutative learning. $F_0X$ converged to the shape transformation. $F_1X$ converged to the color transformation. **b**, A typical example of an ablation study. Both colour and shape transformation were conducted by only by $F_1X$. **c**, Another example of an ablation study. Both single-transformations collapsed. **d** Learning changes in invariance in the control condition. **e, f** Same changes in the ablation condition. Blue, colour invariance of $F_0X$; red, colour invariance of $F_1X$; magenta, shape invariance of $F_0X$; and cyan, shape invariance of $F_1X$. **g**, Median invariances in control and ablation were 0.92 and 0.37; the distributions in the two groups differed significantly (Mann–Whitney U = 177, n = 100, P < $5 \times 10^{-32}$ two-tailed). **h.** Median commutative loss after training in control and ablation were 2.55e-4 and 0.09; the distributions in the two groups also differed significantly (Mann–Whitney U = 8273, n = 100, P < $2 \times 10^{-15}$ two-tailed).

After training, two 32-dimension latent vectors corresponding to the single-transformations were projected to two-dimensional space using principle component analysis for visualization (Fig. 3). In colour qualia space (Fig. 3a), the same colour with different shapes was arranged into clusters. In shape qualia space (Fig. 3b), the same shapes with different colour were arranged into clusters. Furthermore, we observed that the same shape was arranged linearly from the origin in the shape qualia space. Thus, the latent vectors could be divided into two different metric spaces corresponding to colour and shape.

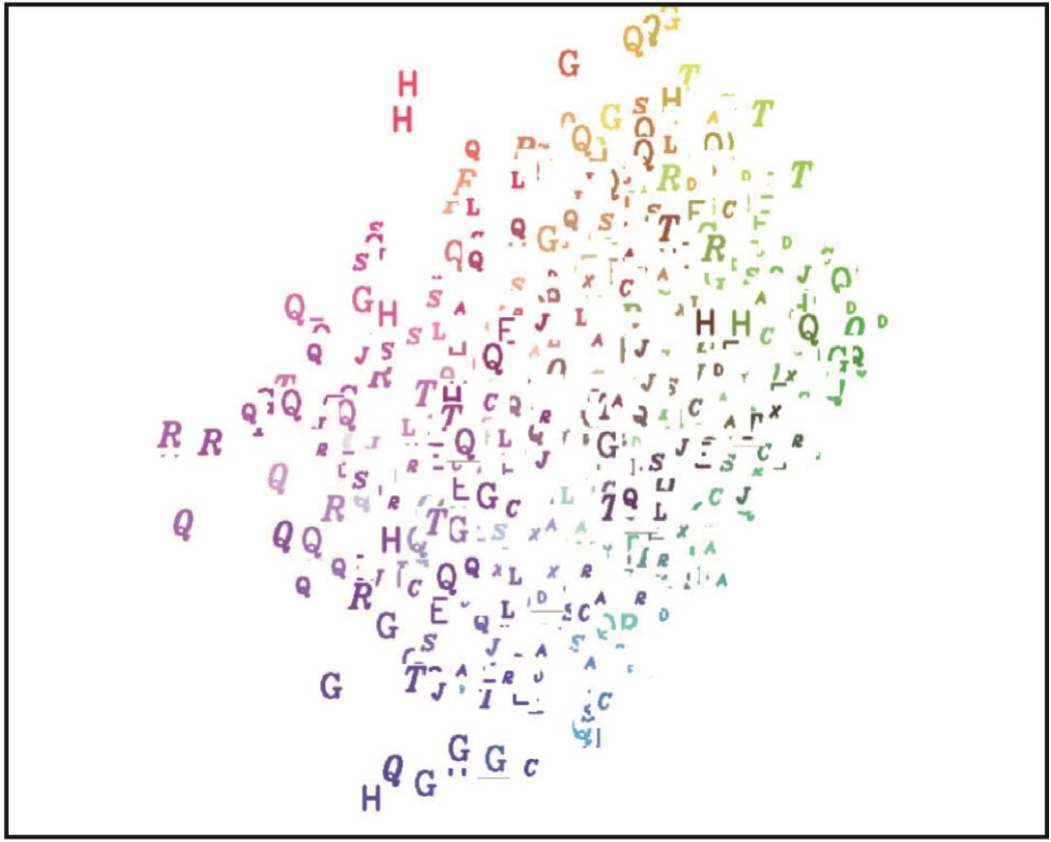

(A)

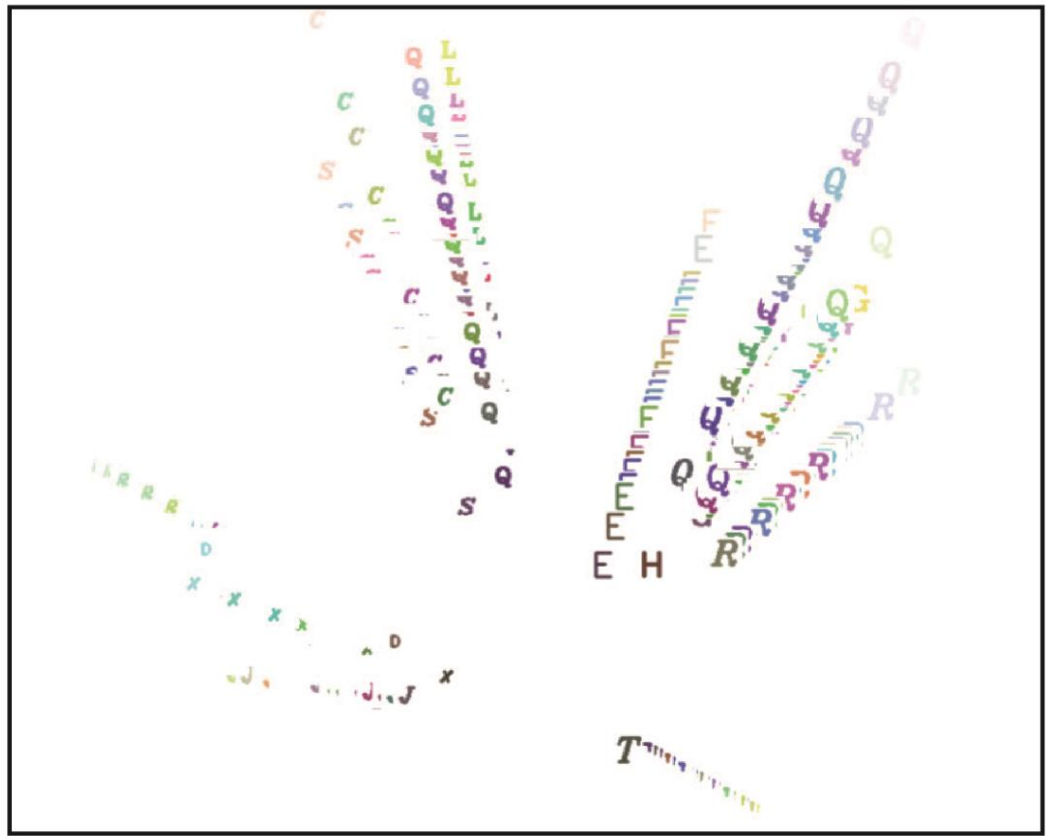

(B)

Figure 3: Learned metric spaces. (a) Metric space corresponding to the colour transformation. (b) Metric space corresponding to the shape transformation. Sixteen different letters in 32 different colours were plotted to learned metric spaces. Each space was mapped in 2D using PCA for visualization. Contribution rate: 100.0% in colour space, 62.2% in shape space. For visualization, the background colour was changed from black to white.

**Discussion**

Here, we showed that the discriminative structure of qualia is a combination of multiple metric spaces that can be constructed autonomously by the algebraic independence learning. First, we established the connection between the algebraic independence between transformations and the generalized invariant transformation equation of a vector transformation parameter. Then, we tested that the brain can autonomously acquire the invariant transformation equation through representation learning to satisfy the algebraic independence between neural network modules and project sensory information onto multiple metric spaces to independently evaluate the differences and similarities in each qualia-type.

Several theories of consciousness (Tononi 2008; Tononi and Edelman 1998; Edelman 2003; Seth *et al.* 2006; Tononi *et al.* 2016) suggest that a fundamental property of conscious states is that they are both integrated (they cannot be divided into independent components) and differentiated (they can choose from an enormous number of conscious states). These theories assumed independence between elements. Integration and independence are therefore contradictory. However, if we assume independence between transformations, integration can be interpreted as the composition by a commutative product, a necessary condition for defining independence. In addition, the independence between elements is interpreted as transformation parameters, each transformation maintains other independent transformation parameters invariant. The differentiation in the conscious property can also be explained by an enormous variation in the composition of the independent transformation parameters. Since groups of neurons responding to invariant properties of objects are more likely to contribute to conscious experience (Tononi and Edelman 1998), the construction of the invariant transformation equations and qualia categorisation should be related to conscious experience.

Sperry (Sperry 1987) argued that consciousness cannot be explained by reductionism alone, which assumes that a whole system can only be reduced to its parts and the law of the parts, because of the existence of downward causation (Sperry 1987; van Gulick 2001; Chalmers 2006; Campbell 1974). Downward causation is a causal relationship from higher level relationships to lower-level parts of themselves. Mind (free will) is an example of the downward causation because it must have a causal power over the brain and body without any additional matter other than body. Learning the algebraic structure used here is also downward causation, because the algebraic structure is a higher-level information structure that cannot be observed directly from its parts, but the information can be evaluated intrinsically. Importantly, algebra is not top-down information processing that requires additional brain regions for calculation. Algebra is macro-relationship rather than a state of micro-elements. Each macro-component consists of micro-elements and requires no additional matter other than the micro-elements. To our knowledge, algebraic structure learning is a new mathematical formulation of the downward causation between the macro-micro hierarchy (Simon 1977) and a mechanism of the downward causation is not addressed in current major theories of consciousness (Seth and Bayne 2022).

Our experiment is limited to the categorisation of two qualia and our formulation assumes that the sensory input has only a single object in a scene. When the number of transformations is greater than two, algebraic independence can be defined between arbitrarily chosen pairs of transformations. Therefore, the variety of transformation pairs satisfying the algebraic independence provides the possibility to define a more complex algebraic structure, including an asymmetry structure or conditional independence (Simpson 2018). With such a complex algebraic structure, extensive research is needed in the future to construct a more realistic qualia structure. In addition, the mathematical principle that determines the conscious access (Baars 2002) from the artificial qualia structure needs to be established. Regarding the single object restriction, the addition operation that is used in the multi-object representation (Christopher *et al.* 2019) satisfies commutativity in the condition of the algebraic independence, suggesting the possibility of integrating the multi-object decomposition into our theory. This issue is also important because it involves the formulation of target selection by attention from the complex multi-object scenes.

**Conclusions**
Our mathematical theory of qualia discrimination and constructive method show that how brain can autonomously construct multiple metric spaces corresponding to qualia types through algebraic structure learning between neural networks without constraining axes to be independent and algebraic structure learning is a new formulation of the downward causation, suggesting that our theory can contribute to the further development of the mathematical theory of consciousness.

*Reference*

1. Müller, J., *Elements of Physiology*. Hard Press (2017).
2. Freemann, W. J., Simulation of Chaotic EEG Patterns with a Dynamic Model of the Olfactory System. *Biol. Cybern.* **56**, 139–150 (1987)
3. Jackendoff, R., *Consciousness and the Computational Mind*. The MIT Press, Cambridge, Massachusetts (1987).
4. O'Regan, J. K. & Noë, A., A sensorimotor account of vision and visual consciousness. *Behav. and Brain Sci.* **24**, 939–1031 (2001).
5. Stanley, R.P., Qualia Space. *J. Conscious. Stud.* **6**, 49–60 (1999).
6. Fekete, T. & Edelman, S., Towards a computational theory of experience. *Conscious. Cogn.* **20**, 807–827 (2011).
7. Yoshimi, J., Phenomenology and connectionism. *Front. Psychol.* **2**, 10.3389/fpsyg.2011.00288 (2011).
8. Prentner, R., Consciousness and topologically structured phenomenal spaces. *Conscious. Cogn.* **70**, 25–38 (2019).
9. Northoff, G., Tsuchiya, N. & Saigo, H., Mathematics and the Brain: A Category Theoretical Approach to Go Beyond the Neural Correlates of Consciousness. *Entropy* **21**, 10.3390/e21121234 (2019).
10. Signorelli, C.M., Wang, Q. & Khan, I., A Compositional Model of Consciousness based on Consciousness-Only. *Entropy* **23**, 10.3390/e23030308 (2021).
11. Chalmers, D.J., *The Conscious Mind*. Oxford University Press, (1996).
12. Seth, A.K. & Bayne, T., Theories of consciousness. *Nat. Rev. Neurosci.* **23**, 439–452 (2022).
13. Del Pin, S.H, Skóra, Z., Sandberg, K. Overgaard, M. & Wierzchoń, M., Comparing theories of consciousness: why it matters and how to do it. *Neurosci. Consciousness* **7**, 1–8 (2021).
14. Browning, H. & Veit, W., The Measurement Problem in Consciousness. *Philos. Top.* **48**, 85–108 (2020).
15. Pinto, Y & Stein. T., The hard problem makes the easy problems hard – a reply to Doerig et al. . *Cong. Neurosci.* **12**, 97–98 (2021).
16. Piaget, J., *The Psychology of Intelligence*. Routledge and Kegan Paul, London (1950).
17. Husserl, H., *The Idea of Phenomenology*. The Hague: M. Nijhoff (1958).
18. Graziano, M.S., A conceptual framework for consciousness. *Proc. Natl. Acad. Sci. USA* **119**, 10.1073/pnas/2116933119 (2022).
19. Searle, J.R., Minds, Brains, and Programs. *Behav. Brain Sci.* **3**, 417–457(1980).
20. Kuniyoshi, Y., Fusing autonomy and sociality via embodied emergence and development of behaviour and cognition from fetal period. *Philos. Trans. R. Soc. B: Biol. Sci.*, 10.1098/rstb.2018.0031 (2018).

21. Chrisley, R., Philosophical foundations of artificial consciousness. *Artif. Intell. Med.* **44**, 119–137 (2008).
22. Clowes, R.W & Seth, A.K., Aximos, properties and criteria: Roles for synthesis in the science of consciousness. *Artif. Intell. Med.* **44**, 91–104 (2008).
23. Seth, A., Explanatory Correlates of Consciousness: Theoretical and Computational Challenges. *Cogn. Comput.* **1**, 50–63 (2009).
24. Fumero, M., Cosmo, L., Melzi, S., Rodolà, E., Learning disentangled representation via product manifold projection, *PMLR* **139** (2021).
25. Prakash C., Fields, C., Hoffman, D.D., Prentner, R. & Singh, M., Fact, Fiction, and Fitness. *Entropy* **22**, 10.3390/e22050514 (2020).
26. Nagel, T., What is it like to be a bat? *The Philosophical Review* **83**, 435–450 (1974).
27. Kleiner, J., Mathematical Models of Consciousness. *Entropy* **22**, 10.3390/e22060609 (2020).
28. Simpson, A., Category-theoretic Structure for Independence and Conditional Independence. *Electron. Notes Theor. Comput. Sci.* **336**, 281–297 (2018).
29. Sperry, R., Structure and Significance of the Consciousness Revolution. *J. Mind Behav.* **8**, 37–66 (1987).
30. Van Gulick, R., Reduction, Emergence and Other Recent Options on the Mind/Body Problem. *J Conscious Stud.* **8**, 1–34 (2001).
31. Chalmers, D. Strong and Weak Emergences, *The Re Emergence of Emergence: The Emergenist Hypothesis from Science to Religion*, eds. Clayton P. and Davies P., Oxford: Oxford University Press, 244-254 (2006).
32. Campbell, D.T., 'Downward Causation' in Hierarchically Organised Biological Systems. *In Studies in the philosophy of biology; Reduction and related problems*, eds. Ayala., F. and Dobzhansky, T., University of California Press, 179–186 (1974).
33. Simon, H.A, *The Organization of Complex System*. In: Models of Discovery. Boston Studies in the Philosophy of Science **54**, Springer, 245-261 (1977).
34. Higgins, I. et al., $\beta$-VAE: Learning Basic Visual Concepts with a Constrained Variational Framework. *Proc. of ICLR* (2017).
35. Comon, P., Independent Component Analysis, A new concept? *Signal Process.* **36**, 287–314 (1994).
36. Pitts, W. & McCulloch, W.S., How We Know Universals: The Perception of Auditory and Visual Forms. *Bull. Math. Biophys.* **9**, 127–147 (1947).
37. Cassirer E., The Concept of Group and the Theory of Perception. *Philos. Phenomenol. Res.* **5**, 1–36 (1944).
38. Hoffman, W. C., The Lie Algebra of Visual Perception. *J. Math. Psychol.* **3**, 65–98 (1966).
39. Otsu, N., Recognition of Shape and Transformation: An Invariant-theoretical foundation. *Sci. on

*Form: Proc. of the 1st Int. Symp. for Sci. on Form*, KTK Scientific Publishers, Tokyo (1986).

40. Higgins, I., et al., Towards a Definition of Disentangled Representations. *arXiv preprint*, arXiv:1812.02230 (2018).
41. Puthawala, M. et al., Globally Injective ReLU Networks. *J. Mach. Learn. Res.* **23**, 1–55 (2022).
42. Krizhevsky, A., Sutskever, I. & Hinton, G.E., ImageNet Classification with Deep Convolutional Neural Networks. In *Adv. In Neural Inf. Process. Syst.* **25**, 1097-1105 (2012).
43. Lin, M., Chen, Q. & Yan, S., Network in Network. *arXiv: 1312.4400* (2014).
44. Kingma, D.P. & Dhariwal, P., Glow: Generative Flow with invertible 1x1 Convolutions. *In Adv. in Neural Inf. Process Syst.* **31**, 10236-10245(2018).
45. Lezcano-Casado, M., Trivializations for Gradient-based Optimization on Manifolds. *In Adv. in Neural Inf. Process Syst.* **32**, 9154-9164(2019).
46. Salimans, T. & Kingma, D.P., Weight Normalization: A Simple Reparametrization to Accelerate Training of Deep Neural Networks. *In Adv. In Neural Inf. Process Syst.* **29**, 901-909 (2016).
47. Hoffer, E., Banner, R., Golan, I. & Soudry, D., Norm matters: efficient and accurate normalization schemes in deep networks. *In Adv. In Neural Inf. Process Syst.* **31**, 2164-2174 (2018).
48. Liu, L. et al., On the Variance of the Adaptive Learning Rate and Beyond. *Proc. of ICLR* (2020).
49. Paszke, A. et al., PyTorch: An Imperative Style, High-Performance Deep Learning Library. *In Adv. in Neural Inf. Process Syst.* **32**, 8024-8035 (2019).
50. Tononi, G., Consciousness as Integrated Information: a Provisional Manifesto. *Biol. Bull.* **215**, 216–242 (2008).
51. Tononi, G. & Edelman, G.M., Consciousness and Complexity. *Science* **282**, 1846–1851 (1998).
52. Edelman, G.M., Naturalizing consciousness: A theoretical framework. *Proc. Natl. Acad. Sci. USA* **100**, 5520–5524 (2003).
53. Seth, A.K., Izhikevich, E., Reeke, G.N. & Edelman, G.M., Theories and measures of consciousness: An extended framework. *Proc. Natl. Acad. Sci. USA* **103**, 10799–10804 (2006).
54. Tononi G., Boly, M., Massimini, M. & Koch, C., Integrated information theory: from consciousness to its physical substrate. *Nat. Rev. Neurosci.* **17**, 450–461, 2016.
55. Baars, B.J., The conscious access hypothesis: origins and recent evidence. *Trends Cogn. Sci.* **6**, 47–52 (2002).
56. Christopher, P. et al., MONet: Unsupervised Scene Decomposition and Representation. *arXiv preprint*, arXiv:1901.11390 (2019).

**Acknowledgements**

This paper is based on results obtained under a Grant-in-Aid for Scientific Research (A) JP22H00528.

**Data availability**

All data needed to evaluate the conclusion in the paper are present in the paper and/or the supplementary materials. Additional data can be requested from the corresponding author.

**Code availability**

Source code are available from the corresponding author on reasonable request.

**Author Contributions**

YO substantially contributed to the study conceptualization and the mathematical theory. WS and YO contributed to the experimental design. WS developed a part of software, neural networks design and dataset. YO mainly developed the software, neural networks design and loss function. YO wrote original draft preparation. YK supervised the project. All authors discussed the results and commented on the manuscript.

**Competing Interests Statement**

The authors declare no conflicts of interest associated with the manuscript.